\documentclass[aps,scriptaddress, showkeys,nofootinbib,reprint]{revtex4-1}
	\usepackage{hyperref}
	\usepackage{cancel} 
	\usepackage{amsmath,amssymb} 
	\usepackage{makeidx} 
	\usepackage{amsfonts}
	\usepackage[ansinew]{inputenc}
	\usepackage[usenames,dvipsnames]{pstricks}    
	\usepackage{subfigure}
	\usepackage{epsfig}
	\usepackage{pst-grad}
	\usepackage{pst-plot} 
	\usepackage{comment}
	\usepackage{pstool}
\makeindex
\newcommand{\beq}{\begin{equation}}
\newcommand{\eeq}{\end{equation}}
 
\newcommand{\bea}{\begin{eqnarray}}
\newcommand{\ea}{\end{eqnarray}}
\newcommand{\barr}{\begin{array}}
\newcommand{\earr}{\end{array}}

\begin{document}

\title{Fine-tuning challenges for the matter bounce scenario}

\author{Aaron M. Levy}
\email{aaronml@princeton.edu}
\affiliation{Department of Physics, Princeton University, Princeton, NJ 08544, USA}

\begin{abstract}
  A bouncing universe with a long period of contraction during which the average density is pressureless (the same equation of state as matter) as cosmologically observable scales exit the Hubble horizon has been proposed as an explanation for producing a nearly scale-invariant spectrum of adiabatic scalar perturbations. A well-known problem with this scenario is that, unless suppressed, the energy density associated with anisotropy grows faster than that of the pressureless matter, so the matter-like phase is unstable.  Previous models introduce an ekpyrotic phase after the matter-like phase to prevent the anisotropy from generating chaotic mixmaster behavior.  In this work, though, we point out that, unless the anisotropy is suppressed first, the matter-like phase will never start and that suppressing the anisotropy requires extraordinary, exponential fine-tuning. 
\end{abstract}

\date{\today}

\maketitle
Matter bounce models were introduced to provide a simple mechanism for generating a  scale-invariant spectrum of adiabatic perturbations in accord with observations \cite{Komatsu:2008hk,Ade:2015lrj,Ade:2015ava, Sievers:2013ica}. The basic idea is that quantum fluctuations naturally generate a scale-invariant spectrum of adiabatic curvature perturbations during a contracting phase if the dominant density is a pressureless (\emph{i.e.,} matter-like) fluid, such as a scalar field rolling along an exponential potential \cite{Wands:1998yp, Finelli:2001sr, Allen:2004vz}.  If the matter-like phase is followed by a nonsingular bounce, say, the scale-invariant spectrum can be preserved after the bounce and provide an explanation of the observed fluctuations in the microwave background and of the large-scale structure of the universe.  This scenario is referred to as ``matter bounce" \cite{Brandenberger:2012zb}. It resolves the horizon  problem (and in some incarnations, the flatness problem) of standard Big Bang cosmology \cite{Brandenberger:2016vhg}, generates the observed perturbations, and avoids the multiverse problem \cite{gibbons1985very, Vilenkin:1983xq, Guth:2000ka, Guth:2013epa, Guth:2013sya,  Linde:2014nna} of inflation  \cite{PhysRevD.23.347,Albrecht:1982wi,Linde:1981mu}. 

One well-known problem with the matter bounce scenario is its overproduction of tensor fluctuations during the matter-like contracting phase. This issue has been studied extensively \cite{Cai:2012va,Quintin:2015rta} with various proposed resolutions \cite{Cai:2011zx, WilsonEwing:2012pu, Cai:2014jla, Fertig:2016czu}. But perhaps the biggest problem for the matter bounce scenario is the instability of the contracting matter-like phase to anisotropy.  If unchecked, anisotropy rapidly dominates the energy budget of the universe-- spoiling the equation of state responsible for the scale invariant spectrum-- and ultimately leading to chaotic Belinskii-Khalatnikov-Lifshitz (BKL) behavior \cite{belinskii1970oscillatory}.

Most works attempting to resolve the anisotropy problem focus on avoiding BKL instability after the matter-like contracting phase and thus after the generation of scale-invariant superhorizon perturbations \cite{Cai:2012va, Cai:2013vm, Quintin:2015rta}. They argue that if anisotropy is subdominant by the end of the matter-like contraction, then an ensuing phase of ekpyrosis will render it negligible thereafter. Suppressing the anisotropy after the matter-like phase is far too late, though, as we emphasize in this paper.  Unless the anisotropy is exponentially suppressed before the matter-like phase begins, it rapidly overtakes the energy density of matter before a sufficient number of scale-invariant modes have been generated. 

In this work, we examine the anisotropy problem and demonstrate that it requires extreme, exponential tuning of the initial conditions-- exponentially more tuning than required to resolve the flatness problem, for example. In Sec.~\ref{qap}, we summarize the anisotropy problem along lines similar to those in Ref.~\cite{Bozza:2009jx}. In Sec.~\ref{new}, we argue that suppressing the anisotropy requires a protracted isotropizing phase \emph{prior} to the matter-like phase,   and moreover that the degrees of freedom responsible for the matter-like phase must be coupled to those driving the isotropizing phase. In Sec.~\ref{cons}, we construct an example of this sort involving a canonical scalar field with a specially constructed potential.  In Sec.~\ref{disc}, we show how resolving the anisotropy problem requires extreme fine-tuning of this potential. In Sec.~\ref{disc2}, we conclude, agruing that this extreme fine-tuning is a generic property of the matter bounce scenario.

In what follows, we employ reduced Planck units and metric signature $(-+++).$ 
\section{Quantifying the Anisotropy Problem}\label{qap}
In this section, we review the anisotropy problem for matter-dominated, contracting universes, demonstrating that the growth of anisotropy is exponentially sensitive to the number of modes that leave the horizon during the matter-like phase. This result is summarized in Eq.~\eqref{matterani}.

In a flat Friedmann-Robertson-Walker universe driven by a stress-energy component, $X$, with a constant equation of state $\epsilon=-\dot H/H^2$,  the scale factor, $a$, is related to the Hubble parameter, $H\equiv \dot a/a$, in the following way
\beq
\label{sf}
\frac{a_f}{a_i}=\left(\frac{H_i}{H_f}\right)^{1/\epsilon},
\eeq
where subscripts $i$ and $f$ denote initial and final values. In the above, overdots denote derivatives with respect to coordinate time, $t$. 
The ratio of the energy density in anisotropy $(\propto a^{-6})$ to that in $X$ $(\propto a^{-2\epsilon})$, $f\equiv \rho_\sigma/\rho_X$, scales as 
\beq
\label{ani}
\frac{f_f}{f_i}=\left(\frac{a_f}{a_i} \right)^{2(\epsilon-3)}=\left(\frac{H_i}{H_f}\right)^{2\left(1-\frac{3}{\epsilon}\right)},
\eeq
where the second equality follows from Eq.~\eqref{sf}. A perturbation with comoving wavenumber will exit the horizon when, $k=a |H|$. As the universe evolves, $a|H|$ grows (for $\epsilon>1$), taking shorter and shorter wavelengths outside the horizon. That is, between times $t_f$ and $t_i$
\beq
\label{efold}
N\equiv\ln\left(\frac{a_fH_f}{a_iH_i}\right) =\left(1-\frac{1}{\epsilon}\right)\ln\left(\frac{H_f}{H_i}\right)
\eeq
 $e$-foldings of scales will have exited the horizon. Combining Eqs.~\eqref{efold} and \eqref{ani} yields
\beq
\label{aniefold}
\frac{f_f}{f_i}=\exp\left( -2N\left(\frac{\epsilon-3}{\epsilon-1}\right)\right).
\eeq
During ekpyrosis, $\epsilon>3,$ so the right side of Eq.~\eqref{aniefold} decreases exponentially with $N$, reflecting the isotropizing power of ekpyrosis.  By contrast, during matter-dominated contraction, $\epsilon=3/2$, so the right side of Eq.~\eqref{aniefold} \emph{grows} exponentially with $N$,
\beq
\label{matterani}
f_f/f_i=e^{6N}.
\eeq
This quantifies the anisotropy problem of matter-dominated contraction. Unless $f_i$ is fantastically small, Eq.~\eqref{matterani} shows that anisotropy will overtake matter after only a few $e$-foldings of scales have left the horizon. 

Past works have simply \emph{assumed} $f_i$ to be small, arguing that if $f_f$ does not exceed unity by the end of the matter-like contraction, then an ensuing phase of ekpyrotic contraction will ensure that it never does.  The problem with the above logic is  in assuming $f_i\sim\mathcal{O}(e^{-360})$ (if 60 $e$-foldings of scales are generated). For comparison, consider the flatness problem of standard Big Bang cosmology. In its most extreme version, wherein radiation-dominated expansion is assumed to begin at or near Planckian energy density, the fractional contribution of spatial curvature, $\Omega_K\propto 1/(aH)^2$, increases by a factor of  $(\Omega_{K})_f/(\Omega_{K})_i=(\dot a_i/\dot a_f)^2=t_i/t_f=(T_f/T_i)^2\sim e^{73}$, where $T$ is temperature, and for simplicity, we have assumed radiation domination all the way to present-day at $T_f=2.7K$. Thus, the anisotropy problem of the matter bounce scenario is many, \emph{many} orders of magnitude worse than the flatness problem of standard Big Bang cosmology. It involves a factor of $e^{-287}$ more tuning. 

\section{The necessity of coupling}\label{new}
Without a powerful isotropizing phase before matter domination, it is clear from the last section that anisotropy quickly spoils the generation of scale-invariant modes.  But suppressing anisotropy before matter domination is impossible unless the degrees of freedom responsible for the isotropizing phase are coupled to those responsible for the matter-like phase, as we now show.

Consider a universe with three components:  anisotropy, pressureless matter, and a third stress-energy component, $X$, which will be used to suppress anisotropy.  Since anisotropy grows faster than matter, suppressing anisotropy with $X$ requires that $X$ grows faster than both anisotropy and matter. For example, $X$ might be an ekpyrotic field. Thereafter, this component, which begins greater than matter and grows faster than matter must somehow give way to matter. If $X$ is decoupled from the matter, such a transition is impossible.  Either $X$ must decay directly into matter, or else it must drive the matter-like phase itself. In either case, suppressing the anisotropy imposes extreme fine-tuning requirements on the Hubble parameter, as we will show below.

The decay scenario suffers additionally from the tight constraint that the decay products must gravitate like nonrelativistic matter and nothing stiffer that might spoil a matter-like background. For example, from the reasoning of the previous section, any relativistic species, produced even in modest amounts, will grow faster than matter by a factor of $\exp(2N)$, quickly spoiling the matter-like phase. Therefore, we will focus on the scenario without decay, presenting one realization in which $X$ is a scalar field whose potential is specially constructed to produce both phases, first (stiff) ekpyrotic- and then (soft) matter-like contraction. We show that suppressing the anisotropy is possible only if the potential is extremely fine-tuned. Thus, the tuning of the anisotropy is traded for a tuned potential, and hence a tuned Hubble parameter.

\section{Stiff-to-soft  Model}\label{cons}
\begin{figure}
\includegraphics[scale=.5]{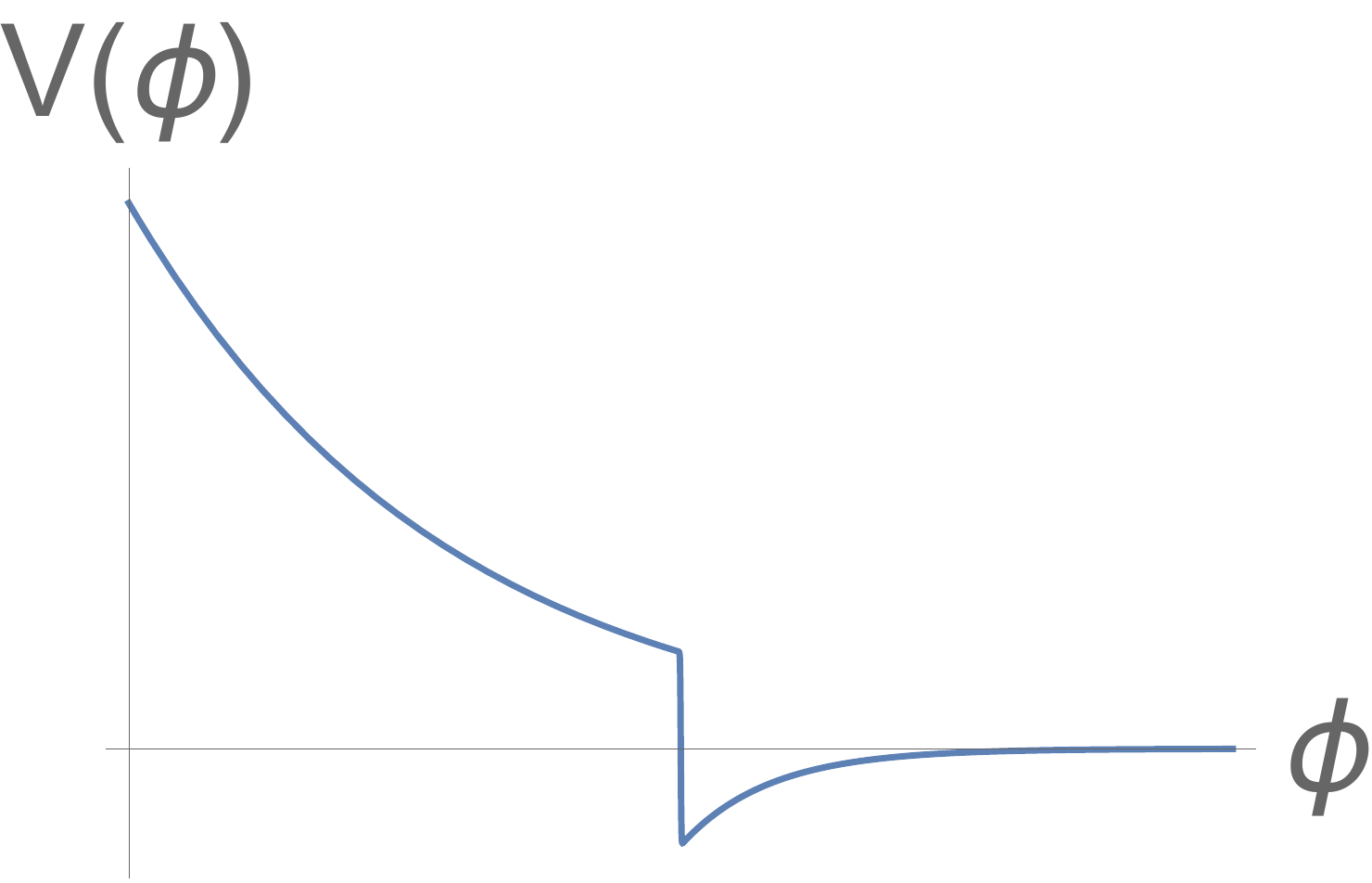}
\caption{\label{v} This shows the potential energy density in Eq.~\eqref{mbpotential}. The field moves from right to left across the figure, transitioning from ekpyrotic to matter-like contraction after the kink. }
\end{figure}

In this section, we present a toy model in which the universe undergoes a phase of ekpyrotic contraction before transitioning into matter-like contraction. The stiff-to-soft transition is possible because both phases are driven by the same scalar field, $\phi$, whose potential energy density, $V(\phi)$, pictured schematically in Fig.~\ref{v}, is specially constructed to obtain this behavior. As $\phi$ moves from the far right of Fig.~\ref{v} to the left, the universe contracts with an ekpyrotic equation of state $\epsilon_{ek}>3$ until it crosses the kink in the middle of the figure. Thereafter, the field runs \emph{up} the potential, driven by Hubble \emph{anti}friction, and the universe contracts with the same equation of state as pressureless matter, $\epsilon_{md}\equiv 3/2$. As discussed, the purpose of the ekpyrotic phase is to suppress the anisotropy so that the succeeding phase remains matter-like and thereby generates a scale-invariant spectrum of adiabatic perturbations. 
\subsection{The equations of motion and the solution}
\begin{figure}
\includegraphics[width=.5\textwidth]{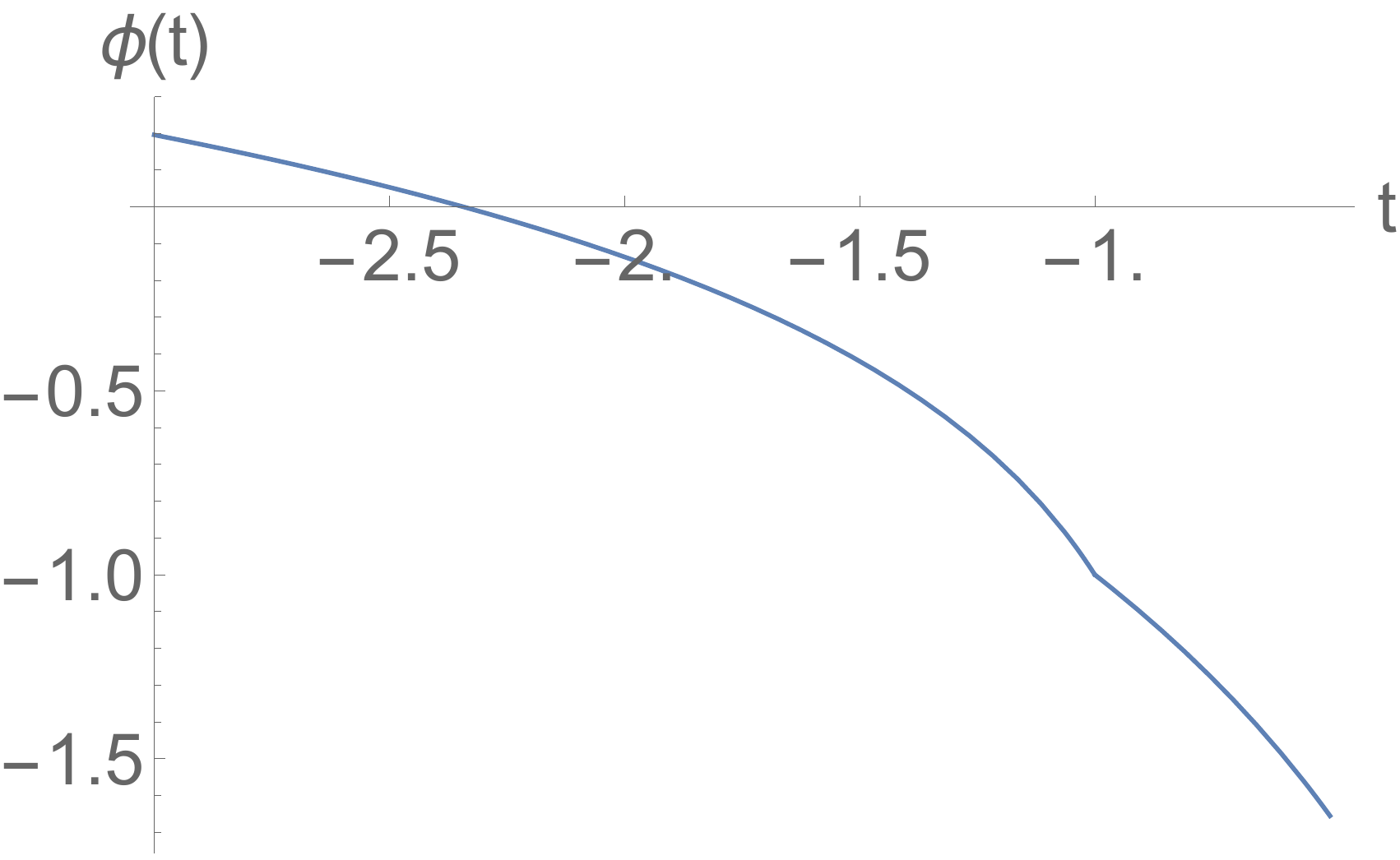}
\caption{\label{phioftnum} This shows a numerical solution to Eq.~\eqref{sfe} superimposed over the analytic result in Eqs.~\eqref{phieksol} and \eqref{phimdsol} (where $\phi_e=t_e=-V_{md}^i=-1$, $\epsilon_{ek}=6$, and $\Delta \phi=10^{-4}$). The two curves are indistinguishable. For $t<t_e$, the solution is ekpyrotic, and for $t>t_e$ it is matter-like.}
\end{figure}

The Lagrangian density is
\beq
\label{mblagrangian}
\mathcal{L}=\frac{1}{2}R-\frac{1}{2}(\partial\phi)^2-V(\phi),
\eeq
where 
\begin{eqnarray}
\label{mbpotential}
V(\phi)&=&V_{md}(\phi)+\Theta\left(\frac{\phi-\phi_e}{\Delta\phi}\right)\times\nonumber\\
&&\left(V_{ek}(\phi)-V_{md}(\phi)\right),\\
V_{ek}(\phi)&\equiv&-V^f_{ek}\exp\left(-\sqrt{2\epsilon_{ek}}(\phi-\phi_{e}) \right),\\
V_{md}(\phi)&\equiv&V^i_{md}\exp\left(-\sqrt{2\epsilon_{md}}(\phi-\phi_{e}) \right),\\
\Theta(\phi)&\equiv&\left(1+\text{tanh} \,\phi\right)/2,
\end{eqnarray}
$V^f_{ek}>0$ is the magnitude of the potential energy density at the end of ekpyrosis, $V^i_{md}>0$ is the magnitude of the potential energy density at the onset of matter domination, $\phi_e$ is the field value at which ekpyrosis transitions into matter domination, and $\Delta \phi$ sets the width of the transition.  We will consider the limit of a rapid transition, namely $\Delta\phi \to 0$, so that the changeover from ekpyrosis to matter domination can be approximated by a Heaviside $\theta$ function, \emph{i.e.,} $\Theta(\frac{\phi-\phi_e}{\Delta\phi})\to\theta(\phi-\phi_e)$. 
In this limit, 
\beq
 V(\phi) \approx
  \begin{cases} 
      \hfill V_{ek}(\phi)   \hfill & \text{ for $\phi>\phi_e$} \\
      \hfill V_{md}(\phi)   \hfill & \text{ for $\phi<\phi_e$},
  \end{cases}
\eeq
so the scalar field generates ekpyrotic contraction to the right of the kink and matter-like contraction to the left of the kink. 
The solution to the equations of motion,
\begin{eqnarray}
H^2&=&\frac{1}{2}\left(\frac{1}{2}\dot\phi^2+V(\phi)\right), \label{fman}\\
0&=&\ddot\phi+3H\dot\phi+V_{,\phi},\label{sfe}
\end{eqnarray}
is given by
\begin{eqnarray}
a_{ek}(t)&=&a_e\left(1+(t_e-t)\sqrt{\frac{V_{ek}^f}{\epsilon_{ek}-3}}\epsilon_{ek}\right)^{1/\epsilon_{ek}}\label{aeksol}\\
\phi_{ek}(t)&=&\phi_e+\sqrt{2\epsilon_{ek}}\ln \left(\frac{a_{ek}(t)}{a_e}\right)\label{phieksol}
\end{eqnarray}
for $t<t_e$ and by
\begin{eqnarray}
a_{md}(t)&=&a_e\left(1+(t_e-t)\sqrt{\frac{V_{md}^i}{3-\epsilon_{md}}}\epsilon_{md}\right)^{1/\epsilon_{md}}\label{amdsol}\\
\phi_{md}(t)&=&\phi_e+\sqrt{2\epsilon_{md}}\ln \left(\frac{a_{md}(t)}{a_e}\right)\label{phimdsol},
\end{eqnarray}
for $t>t_e$, where $t_e$ is the time at which $\phi=\phi_e,$ $a_e\equiv a(t_e),$ and  $V^i_{md}=3V^f_{ek}/(2(\epsilon_{ek}-3))$.  Figure~\ref{phioftnum} shows excellent agreement between this analytic solution and a numerical solution to the equations of motion. Since the rest of this section is devoted to a derivation of this solution, the casual reader may skip to Sec.~\ref{disc} with no loss of continuity.  
\subsection{Analytical Derivation}

\begin{figure}
\subfigure[]{
\includegraphics[width=.4\textwidth]{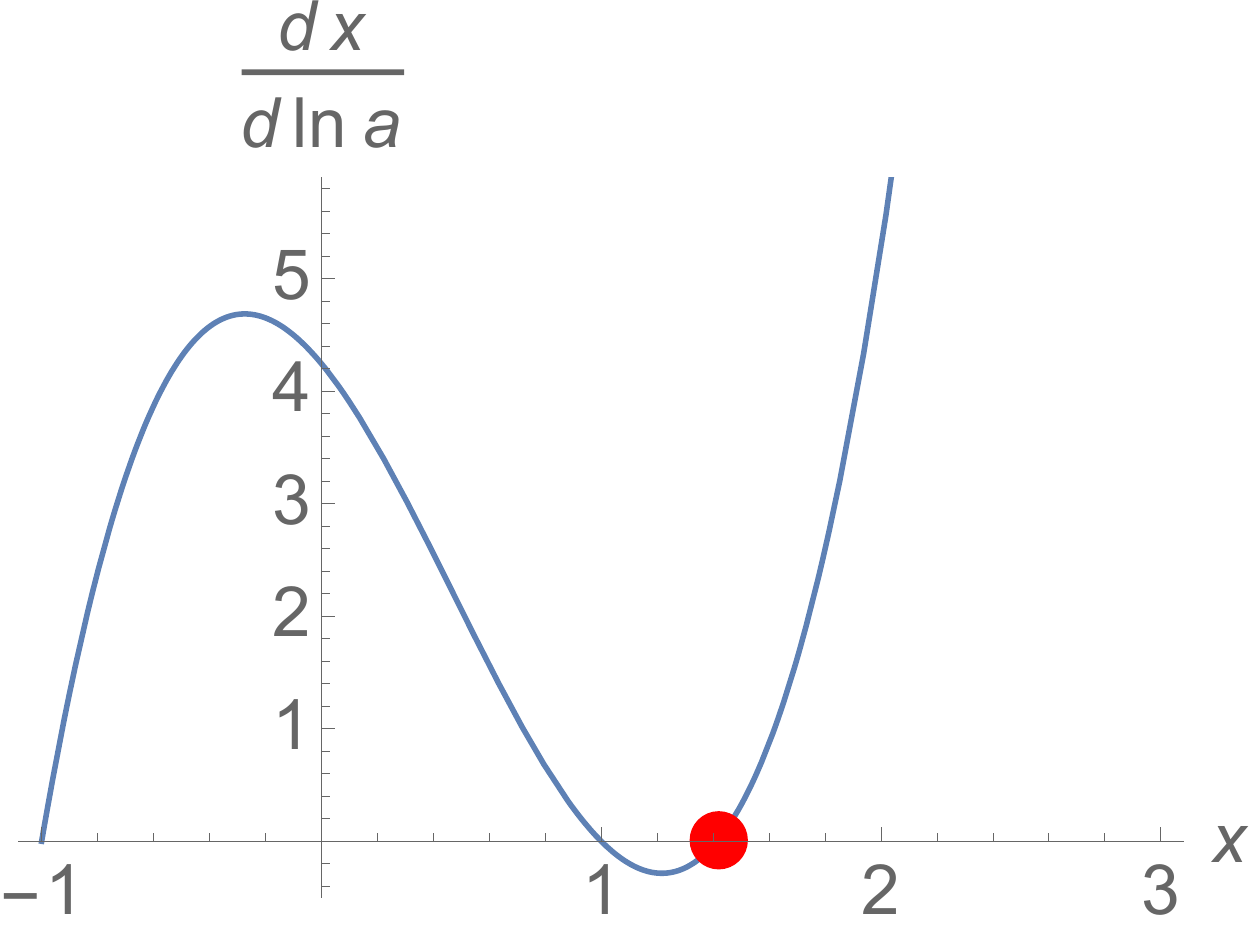}}
\subfigure[]{
\includegraphics[width=.4\textwidth]{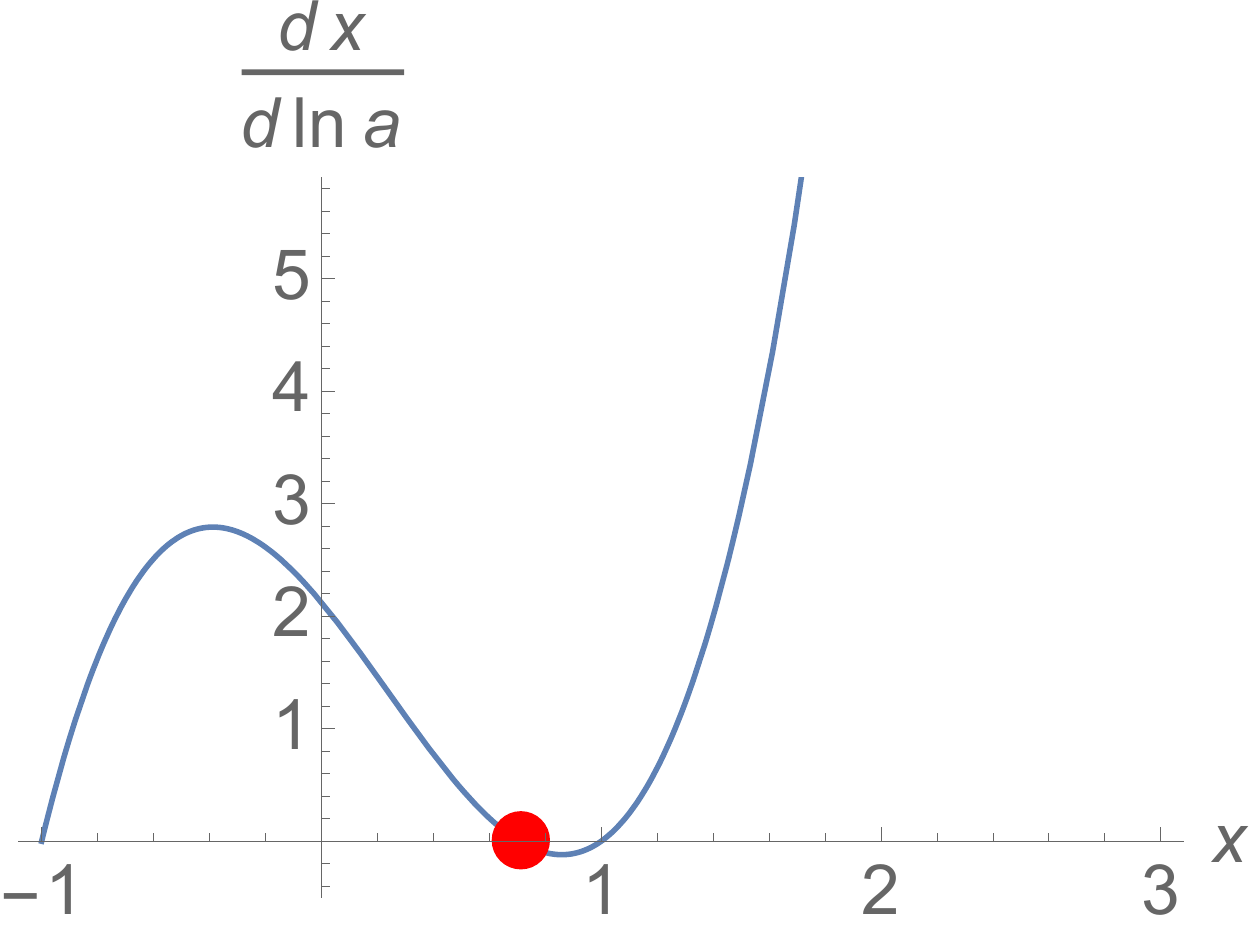}}
\caption{\label{rhseom}This shows the right side of Eq.~\eqref{eomconsteos} during (a) ekpyrosis (for the choice $\epsilon=6$) and (b) matter domination ($\epsilon~=~3/2$). In both cases, the fixed-point solution $x=\sqrt{\epsilon/3}$ is indicated by a red dot. In (a), the blue curve slopes upward at the fixed point, indicating stability. In (b), the blue curve slopes downward at the fixed point, indicating instability.}
\end{figure}

We can gain insight into these dynamics by analyzing the equations of motion in the dimensionless ``$\Omega$-variables'' (or more properly their square roots) $(x,y)\equiv( \frac{\dot\phi}{\sqrt{6}H}, -\frac{\sqrt{|V|}}{\sqrt{3}H})$, characterizing respectively the fractional kinetic and potential energy density in the $\phi$ field. In these variables, the Friedmann equation, Eq.~\eqref{fman}, takes the simple form $y=\sqrt{\pm(x^2-1)}$, where the upper sign corresponds to the case $V<0$ as in the ekpyrotic phase and the lower sign corresponds to $V>0$ as in the matter-like phase. Thus, during ekpyrosis, $x>1$, and during matter domination, $x<1$. In either case, the scalar field equation, Eq.~\eqref{sfe}, can be rewritten as
\beq
\label{eomconsteos}
\frac{dx}{d\ln a}=3(x^2-1)\left(x-\sqrt{\frac{\epsilon}{3}}\right).
\eeq
The function on the right side of Eq.~\eqref{eomconsteos} is plotted in Fig.~\ref{rhseom}. 
There is a fixed-point, scaling solution at $x=\sqrt{\epsilon/3}$.  At first, when $\epsilon=\epsilon_{ek}>3$, this solution corresponds to red dot in Fig.~\ref{rhseom}(a). If the transition is rapid (which we can ensure by taking $\Delta \phi$ small), then the function plotted in Fig.~\ref{rhseom}(a), which applies during the ekpyrotic phase, changes rapidly into that shown in Fig.~\ref{rhseom}(b), which applies during the matter-dominated phase. Since in the matter-dominated phase, $\epsilon=\epsilon_{md}=3/2$, the fixed-point solution corresponding to the red dot is now at $x=1/\sqrt{2}$. As explained in Fig.~\ref{rhseom}, the fixed-point solution is an attractor during ekpyrosis and a repeller during matter domination. Thus, to ensure that matter domination lasts long enough to generate 60 $e$-foldings of scale-invariant modes, the transition must leave the system very close to $x=1/\sqrt{2}$. We now show how to achieve this.

Recall that $t_e$ is the time at which $\phi$ crosses the kink in the potential separating ekpyrosis from matter domination. To find the matching conditions at $t_e$ for the solutions in the two regimes, we first multiply Eq.~\eqref{sfe} by $\dot\phi$ to obtain
\beq
\frac{d}{dt}\left(\frac{1}{2}\dot\phi^2+V\right)=-3H\dot\phi^2,
\eeq
where the last term on the left side follows from the identity $\dot V=V_{,\phi}\dot\phi$. Now we integrate over time from $t_e-\delta$ to $t_e+\delta$ and take the limit $\delta \to 0$. Assuming the right side is finite (though possibly discontinuous) in some neighborhood of the kink, this yields  ``conservation of energy'' across the kink, \emph{i.e.,}
\beq      
\Delta (\dot\phi^2)=-2\Delta (V),
\eeq
where $\Delta (F)\equiv F(t_e^+)-F(t_e^-)$ for any function $F(t)$. Note that this implies continuity of $\phi$ and $H$ across the kink. There is, however, a discontinuity in $\dot\phi$:  the kinetic energy of the field is reduced by the height of the kink. 

Therefore, to ensure that the transition carries the ekpyrotic solution at $x_{before}=\sqrt{\epsilon_{ek}/3}$ to the matter-like solution at $x_{after}=\sqrt{\epsilon_{md}/3}$, the height, $\Delta V$, of the kink must be chosen to satisfy
\beq
\frac{1}{\sqrt{2}}\overset{!}{=}x_{after}=\frac{\dot\phi_{after}}{\sqrt{6}H^f_{ek}}=\sqrt{x_{before}^2-\frac{\Delta V}{3(H^f_{ek})^2}}.
\eeq
Since $(H^f_{ek})^2=V^f_{ek}/(3(x_{before}^2-1))$ and $x_{before}=\sqrt{\epsilon_{ek}/3}$, this gives $\Delta V=(1+\frac{3}{2(\epsilon_{ek}-3)})V^f_{ek}$ or
\beq
\label{match}
V^i_{md}=\frac{3}{2(\epsilon_{ek}-3)}V^f_{ek},
\eeq
as claimed below Eq.~\eqref{phimdsol}.
\section{Analysis of fine-tuning}\label{disc}
We have constructed a cosmological model in which soft (pressureless) matter  overtakes stiff (ekpyrotic) matter. This is necessary for the matter bounce scenario to explain the initial smallness of the anisotropy at the onset of the matter-like phase. Unfortunately, as we will now show, small anisotropy requires an extremely fine-tuned potential. 

First, note that generating $N_{md}$ $e$-foldings of scales during the matter-like  phase immediately requires 
\beq
\label{mdpotwidth}
V^f_{md}/V^i_{md}=\exp\left(6 N_{md}\right).
\eeq
or equivalently, that $|H|$ must grow during the matter-like phase by a factor $\exp\left(3N_{md}\right)$. Although we have modeled the pressureless matter as a scalar field, it is clear from Eq.~\eqref{efold} that this growth in Hubble is independent of the nature of the pressureless matter (so long as it can support density fluctuations, \emph{e.g.}, a scalar field with the same equation of state as matter). During this period, recall from Eq.~\eqref{matterani} that the fractional energy density in anisotropy will have grown by a factor of $\exp\left(6N_{md}\right)$. Therefore, the preceding ekpyrotic phase must suppress anisotropy by at least this much. This requires 
\beq
V^f_{ek}/V^i_{ek}>\exp\left(\frac{6N_{md}}{1-\frac{3}{\epsilon_{ek}}}\right),\label{eklength}
\eeq
or equivalently that $|H|$ must grow by a factor of at least $\exp\left(3N_{md}\right)$ during the ekpyrotic phase. Therefore, combining Eqs.~\eqref{mdpotwidth} and \eqref{eklength} with Eq.~\eqref{match},  we find that from the beginning of the ekpyrotic phase to the end of the matter-like phase, the potential must grow by many orders of magnitude such that
\beq
V^f_{md}/V^i_{ek}>\frac{3}{2(\epsilon_{ek}-3)}\exp\left(\left(12+\frac{18}{\epsilon_{ek}-3}\right)N_{md}\right),
\eeq
or equivalently, that $|H|$ must grow by a factor $\exp\left(6N_{md}\right)$. This is independent of the nature of the degrees of freedom driving ekpyrosis. Thus, we have shown that resolving the anisotropy problem requires an extremely fine-tuned potential. 
\section{Discussion}\label{disc2}
In this work, we have emphasized some of the difficulties imposed by the anisotropy problem on the matter bounce scenario. Collecting the model-independent observations of the previous section, these are:
\begin{enumerate}
\item Generating $N_{md}$ $e$-foldings scale-invariant modes with matter-like contraction requires that $H$ change by $\approx 2.6 N_{md}$ orders of magnitude.  During such a phase, the energy density in anisotropy grows by twice as many orders of magnitude,  $\approx 5.2 N_{md}$. 
\item Therefore, without a powerful suppression mechanism \emph{before} the matter-like contraction, anisotropy rapidly overtakes the matter, thereby spoiling the background required for the scale-invariant spectrum. Previous arguments invoking an isotropizing phase \emph{after} the anisotropy has already grown by this exponentially large factor are analogous to invoking present-day dark energy as a resolution to the flatness problem of standard Big Bang cosmology. 
\item If ekpyrosis is that suppression mechanism, any implementation will require another phase during which $H$ changes by another $2.6N_{md}$ orders of magnitude. It will also require that soft, pressureless matter somehow overtake stiff, ekpyrotic matter. This is impossible unless the degrees of freedom responsible for ekpyrotic contraction are coupled somehow to the pressureless matter. (In the toy model presented here, in which both the stiff and the soft phases result from the same scalar field, this requires fine-tuning a potential over $5.2N_{md}$ orders of magnitude with a kink of just the right height in between.  The only other possibility is to arrange for direct decay of an ekpyrotic field into pressureless matter, which introduces the additional problems discussed in Sec.~\ref{new}.)
\end{enumerate} 

Other attempted resolutions to the problem of small initial anisotropy, involving high-energy, nonlinear modifications to the gravitational action \cite{Middleton:2008rh} or to the equation of state of matter after the matter-like phase \cite{Bozza:2009jx}, suffer from the same fine-tuning constraint discussed in Sec.~\ref{qap}:  it is too late; anisotropy must be exponentially suppressed at the onset of the matter-dominated phase. This was appreciated in Ref.~\cite{Bozza:2009jx}. 

All of these difficulties for the matter bounce scenario are in addition to those discussed in the Introduction, for which plausible resolutions exist, associated with the overproduction of tensor perturbations and the realization of a nonsingular bounce. Any self-consistent implementation of the matter bounce scenario must address all of these difficulties.

We thank L. Berezhiani and G. Tarnopolskiy for many useful discussions. We especially thank R. Brandenberger and P. Steinhardt for their careful reading of the manuscript and useful suggestions to improve it. This research was partially supported by the U.S. Department of Energy under grant number DE-FG02-91ER40671.
\bibliography{bibliography}
\bibstyle{natbib} 
\end{document}